\documentclass[sigconf]{acmart}

\AtBeginDocument{%
  }

\setcopyright{rightsretained}
\acmJournal{PACMMOD}
\acmYear{2025} \acmVolume{3}
\acmNumber{1 (SIGMOD)} \acmArticle{3}
\acmMonth{2} \acmPrice{} \acmDOI{10.1145/3709653}

\usepackage{fontawesome5}
\usepackage{hyperref}

\makeatletter
\gdef\@copyrightpermission{
  \begin{minipage}{0.2\columnwidth}
   \href{https://creativecommons.org/licenses/by/4.0/}{\includegraphics[width=0.90\textwidth]{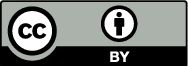}}
  \end{minipage}\hfill
  \begin{minipage}{0.8\columnwidth}
   \href{https://creativecommons.org/licenses/by/4.0/}{This work is licensed under a Creative Commons Attribution International 4.0 License.}
  \end{minipage}
  \vspace{5pt}
}
\makeatother

\usepackage{url} 
\usepackage[utf8]{inputenc} 
\usepackage{booktabs} 
\usepackage{graphicx}
\usepackage{pgfplots}
\usepackage{subcaption}
\usepackage[labelformat=simple]{subcaption}

\pgfplotsset{compat=1.17}

\usepackage{makecell}

\usepackage{listings}
\usepackage{color}

\usepackage{csquotes}

\usepackage[htt]{hyphenat}

\usepackage{pdfpages}

\usepackage{svg}

\usepackage[shortlabels]{enumitem}

\definecolor{dkgreen}{rgb}{0,0.6,0}
\definecolor{gray}{rgb}{0.5,0.5,0.5}
\definecolor{mauve}{rgb}{0.58,0,0.82}
\definecolor{aureolin}{rgb}{0.99, 0.93, 0.0}
\definecolor{bananayellow}{rgb}{1.0, 0.88, 0.21}
\definecolor{canaryyellow}{rgb}{1.0, 0.94, 0.0}
\definecolor{daffodil}{rgb}{1.0, 1.0, 0.19}
\definecolor{electricyellow}{rgb}{1.0, 1.0, 0.0}

\lstdefinestyle{myJava}{ 
  frame=tb,
  language=Java,
  numbers = left,
  stepnumber=1,
  showstringspaces=false,
  columns=flexible,
  basicstyle={\footnotesize\ttfamily},
  numberstyle=\tiny\color{gray},
  keywordstyle=\color{blue},
  commentstyle=\color{dkgreen},
  stringstyle=\color{mauve},
  breaklines=true,
  breakatwhitespace=true,
  tabsize=3,
  morekeywords={var,:,String,Mono,Optional,Future,List,JsonObject,Map,Exception},
  keywordsprefix={@},
  numbers=none
}

\definecolor{bluekeywords}{rgb}{0.13,0.13,1}
\definecolor{greencomments}{rgb}{0,0.5,0}
\definecolor{redstrings}{rgb}{0.9,0,0}

\lstdefinestyle{mySharpC}{
  frame=tb,
  language=[Sharp]C,
  numbers = left,
  stepnumber=1,
  showstringspaces=false,
  columns=flexible,
  basicstyle={\footnotesize\ttfamily},
  keywordstyle=\color{bluekeywords},
  commentstyle=\color{greencomments},
  stringstyle=\color{redstrings},
  breaklines=true,
  breakatwhitespace=true,
  tabsize=3,
  morekeywords={async,await,var,Exception}, 
  numbers=none
}

\usepackage{listings}
\usepackage{color}
\definecolor{dkgreen}{rgb}{0,0.6,0}
\definecolor{gray}{rgb}{0.5,0.5,0.5}
\definecolor{mauve}{rgb}{0.58,0,0.82}
\usepackage{tabularx}
\usepackage{boxedminipage}
\newcolumntype{Y}{>{\centering\arraybackslash}X}
\newcolumntype{M}{>{\centering\arraybackslash}m}
\newcolumntype{S}{>{\centering\arraybackslash}p{0.085\textwidth}}
\newcolumntype{C}{>{\centering\arraybackslash}c}

\usepackage{multirow}

\newcommand{\floor}[1]{\left\lfloor #1 \right\rfloor}

\begin{document}

\title{Online Marketplace: A Benchmark for Data Management in Microservices}

\author{Rodrigo Laigner}
\orcid{0000-0003-2771-7477}
\affiliation{
  \institution{University of Copenhagen}
  \city{Copenhagen}
  \country{Denmark}
}
\email{rnl@di.ku.dk}

\author{Zhexiang Zhang}
\orcid{0009-0001-8493-5254}
\affiliation{
  \institution{University of Copenhagen}
  \city{Copenhagen}
  \country{Denmark}
}
\email{gjm699@alumni.ku.dk}

\author{Yijian Liu}
\orcid{0000-0001-9423-2600}
\affiliation{
  \institution{University of Copenhagen}
  \city{Copenhagen}
  \country{Denmark}
}
\email{liu@di.ku.dk}

\author{Leonardo Freitas Gomes}
\orcid{0009-0004-0875-7320}
\affiliation{
  \institution{Amadeus}
  \city{Nice}
  \country{France}
}
\email{leonardo.gomes@amadeus.com}

\author{Yongluan Zhou}
\orcid{0000-0002-7578-8117}
\affiliation{%
  \institution{University of Copenhagen}
  \city{Copenhagen}
  \country{Denmark}
}
\email{zhou@di.ku.dk}

\renewcommand{\shortauthors}{Rodrigo Nunes Laigner, Zhexiang Zhang, Yijian Liu, Leonardo Freitas Gomes, \& Yongluan Zhou}

\begin{abstract}
Microservice architectures have become a popular approach for designing scalable distributed applications. Despite their extensive use in industrial settings for over a decade, there is limited understanding of the data management challenges that arise in these applications. Consequently, it has been difficult to advance data system technologies that effectively support microservice applications. To fill this gap, we present \textit{Online Marketplace}, a microservice benchmark that highlights core data management challenges that existing benchmarks fail to address. These challenges include transaction processing, query processing, event processing, constraint enforcement, and data replication. We have defined criteria for various data management issues to enable proper comparison across data systems and platforms. 
Through case studies with state-of-the-art data platforms, we discuss the issues encountered while implementing and meeting \textit{Online Marketplace}'s criteria. By capturing the overhead of meeting the key data management requirements that are overlooked by existing benchmarks, we gain actionable insights into the experimental platforms. This highlights the significance of \textit{Online Marketplace} in advancing future data systems to meet the needs of microservice practitioners.
\end{abstract}



\maketitle

\section{Introduction}

Microservice architecture has emerged in the last decade as a popular architectural style in industry settings. This style promotes the decomposition of an application into independent microservices with associated private states. 
From an organizational point of view, these principles allow different teams to manage and evolve their own modules independently. At the same time, it enables new modules to be introduced and deprecated modules to be removed without impacting the application as a whole.
From a technological point of view, each module can be independently deployed on distributed computational resources, allowing for failure isolation and high availability. 
Meanwhile, the message-based communication paradigm serves as a powerful abstraction for triggering tasks in remote microservices, facilitating data replication among microservices, and enabling failure recovery by replaying past events~\cite{olep}.

This industrial popularity has prompted cloud providers to offer rich features targeting microservice deployments~\cite{ms_alibaba,ms_ms,ms_google,ms_aws}, such as specialized container-based technologies to deploy and scale microservices, message brokers to allow for loose-coupled microservices, multi-tenant DBMS technologies to meet the isolation properties of microservices, and application frameworks and side-car technologies to speed up deployment, code evolution and maintenance of microservices.


Despite the benefits of the decoupled design, a recent study~\cite{vldb2021} demonstrates that practitioners encounter several challenges when trying to meet data management requirements in this architecture, including distributed transaction processing, data replication, consistent data and event querying and processing, and enforcing data integrity constraints. In short, microservices are designed to function independently, but in practice, they often rely on each other's data and functionality to complete a workflow. As a result, it is essential to benchmark microservices in a way that accurately reflects the needs of practitioners. Unfortunately, existing microservice benchmarks do not fully capture these real-world requirements~\cite{vldb2021, deathstar, train_tse}. For example, DeathStarBench~\cite{deathstar} and TrainTicket~\cite{train_ticket} do not consider event processing, nor do they specify what data invariants and transactional guarantees are necessary. The absence of a comprehensive benchmark for measuring the performance of data management tasks in microservices leads to the development of custom benchmarks that fail to reflect the complexities of data management in real-world microservices~\cite{netherite}.

\newcommand{\OM}{\textit{Online Marketplace}}


To bridge this gap, we propose \OM, a novel microservice benchmark containing eight microservice types, ten event types, seven query types, including read-only and read-write queries, that reflect the key data management tasks pursued by practitioners, such as distributed transaction processing, data replication, consistent queries, event processing, and data constraint enforcement. To reflect the data management challenges mentioned above, we prescribe seven data management criteria that a data management platform should meet, including functional decomposition, resource isolation, data consistency, and data integrity of microservices. These criteria are meant to embrace the complex nature of deployments found in industry settings and facilitate conducting a fair comparison between different systems on the same basis. To our knowledge, \OM\ is the first microservice benchmark that embraces core data management requirements sought by microservice practitioners.

Based on the definition of \OM, we further developed a data generator and a benchmark driver. The latter can continuously generate and submit transactions to an implementation of \OM. A challenge of implementing the driver is generating transactions at runtime that coherently reflect the dynamic application state, for example, the latest product prices and product versions. Querying the runtime application state imposes an unnecessary and prohibitively expensive workload on the system. To address this problem, we developed a stateful driver, which manages a consistent mirror of some application data and generates coherent transaction inputs. 



We verify the applicability of our benchmark by implementing five versions of \OM\ in three state-of-the-art data platforms, Orleans, Statefun, and a composite solution. Orleans and Statefun are designed for event-driven, distributed stateful applications. The composite solution reflects a usual architecture seek in practice~\cite{vldb2021,olep}. While implementing \OM\ on these platforms, we encountered several limitations that prevented us from fulfilling some of the data management functionalities and criteria sought in practice. We conducted experiments to measure how these competing platforms perform under different workloads and observed their design and architectural implications in performance. Our results show that our benchmark can effectively stress the performance of the platforms and reveal performance and functionality issues. Meanwhile, through \OM, we derive actionable insights to foment the design of futuristic data systems that will meet the expectations of microservice practitioners, who are an important part of the database user community.

By providing a concrete example for database researchers to understand the unmet needs of data management in microservices, the benchmark serves as a testbed for experimenting with novel algorithms, approaches, techniques, and programming models. For instance, a recent study~\cite{smsa} aims to fulfill the data management criteria outlined here to enhance state management features in actor systems for cloud-native applications.
Furthermore, Online Marketplace is a benchmark that not only benefits the real-world microservice applications that served as inspiration for our design~\cite{uber_microservice,uber_proxy,uber_money,uber_payment,nubank,nubank-arc,netflix_eda,netflix_delta,airbnb_integrity,airbnb_double,wix_pitfalls,wix,jet,b2w_query,ifood,podium,synapse} but can also extend beyond existing microservices applications. For instance, besides microservices,
there are trends of other architectural styles and programming models offering trade-offs between complexities in data management and the degrees of coupling. These include modular monolithic architectures, FaaS, Actors, etc. 
As demonstrated in our case studies ($\S$~\ref{sec:platforms}), \OM\ is well positioned to serve as a testbed for these trending cloud application architectures and programming models. The artifacts used in this work are available online~\cite{driver} for reproducibility and extension.

\section{Background}
\label{sec:background}


\noindent\textbf{Data Management Challenges.} 
Despite more than a decade-long employment in the industry, little work in data management research has incorporated the challenges brought about by data-intensive applications following the microservices architectural style. Based on the findings of a recent study that empirically investigated the data management challenges in microservices~\cite{vldb2021} and public reports~\cite{uber_microservice,uber_proxy,uber_money,uber_payment,nubank,nubank-arc,netflix_eda,netflix_delta,airbnb_integrity,airbnb_double,wix_pitfalls,wix,jet,b2w_query,ifood,podium,synapse} of companies with large-scale deployments, we summarize the core data management challenges as follows:


\begin{enumerate}[(i),wide, nosep]
\item \textbf{Ensuring all-or-nothing atomicity.} The asynchronous and non-blocking nature of messages and the lack of interoperability across different data stores make distributed commit protocols difficult to implement, leading developers to either encode their own or eschew the use of synchronization mechanisms.

\item \textbf{Implementing efficient and consistent data processing.} While data is scattered across microservices, some workloads require querying and joining data belonging to different microservices. Therefore, developers often need to implement querying functionalities at the application layer that should belong in the database layer.

\item \textbf{Ensuring data replication correctness.} In order to reduce the expenses associated with querying data from remote microservices, developers often resort to caching or replicating data by subscribing to events generated by other microservices. However, as these events can arrive in any order, developers face challenges in maintaining consistent replication.

\item \textbf{Enforcing cross-microservice data integrity constraints.} As the application is functionally partitioned, data integrity constraints can span multiple microservices. This creates major challenges in constraint enforcement. 

\item \textbf{Ensuring correct event processing order.} 
Developers leverage application-generated events to implement event-based microservice workflows. However, guaranteeing the correct processing order can be challenging due to the asynchronous nature of events. Events can arrive out of order, late, or even duplicated. Such scenarios pose a significant issue when the application logic is sensitive to the processing order of events.

\end{enumerate}




\begin{table}[tb]
\centering
\caption{Comparison of microservice benchmarks}
\vspace{-3ex}
\small
\begin{tabularx}{\columnwidth}{|X|X|X|X|}
\hline
\textbf{Requirement} & \textbf{Criteria} & \textbf{DeathStar \& TrainTicket} & \textbf{Online Marketplace} \\
\hline
Functional Decomposition & Isolation of Resources & Yes & Yes  \\
\hline
Asynchronous Events & Event Processing Order & No & Yes  \\
\cline{1-1} \cline{3-4}
 Event Processing & and Delivery Guarantees & No & Yes  \\
\hline
Distributed Transactions
 & All-or-nothing Atomicity; Isolation Levels Allowed & No & Yes  \\
 \hline
Data Invariants
 & Data Invariant Enforcement & No & Yes \\
 \hline
Data Replication
 & Data Caching and Replication Consistency
 & No & Yes \\
 \hline
 Query Processing & Query Processing Consistency & No & Yes \\
  \hline
\end{tabularx}
\vspace{-1ex}
\label{tab:comparison}
\vspace{-3ex}
\end{table}

\noindent\textbf{Problems of Existing Benchmarks.}
Existing microservice benchmarks like DeathStarBench~\cite{deathstar} and TrainTicket~\cite{train_ticket} do not fully capture these real-world challenges, as shown in Table~\ref{tab:comparison}. DeathStar ~\cite{deathstar} was created to investigate the effects of microservice architectures on hardware and software in system stacks, particularly network and operating systems. TrainTicket focuses on replicating industrial faults in microservices, supporting researchers on investigating fault analysis and debugging. However, they do not take into consideration asynchronous events; thus, challenges related to correct event processing are missing. Furthermore, they do not consider any requirements on data invariants, transactional guarantees, data replication or data querying. Therefore, they target problems that are either oblivious to data management or position data management as not a primary concern.

Similar to microservices, Object-Oriented DBMSs (OODBMS) also adopt the modularization principle, such as encapsulating state and operations into objects within the database. Although, in principle, a microservice can be mapped to an object in OODBMS, we find that OODBMS benchmarks cannot meet the needs that we target. Benchmarks like Cattell, HyperModel, OO7, Justitia, and OCB aim for modeling object relationships in the context of engineering applications and evaluating object clustering algorithms~\cite{DarmontS02}. All these fail to capture the data management requirements in Table~\ref{tab:comparison}.

TPC-W~\cite{tpcw} is a transactional benchmark that models the core aspects of user experience on an e-commerce website, such as browsing pages, checking out books, and searching for keywords. 
TPC-C~\cite{tpcc} was designed to reflect the transaction processing of a wholesale supplier. YCSB~\cite{ycsb} models transactional workloads in the cloud that are not necessarily executed under ACID semantics. All of them assume a traditional monolithic application architecture. Our benchmark models a modern microservice-oriented application and the complex interplay of their components through events, differing substantially from these benchmarks regarding architectural style and data management requirements.

Unibench~\cite{unibench} offers OLTP and OLAP workloads for multi-model databases. However, it does not model the decomposition of microservices, which results in the absence of distributed and encapsulated states as found in microservice applications. On the other hand, stream processing benchmarks like Linear Road~\cite{linear_road} focus on modeling continuous and historical queries, but they do not include transactional workloads.

Therefore, existing benchmarks cannot be used to benchmark the overhead of addressing the correlated data management challenges, which includes the overhead of all-or-nothing atomicity, concurrency control, consistent query processing, consistent data replication, and correct event processing. Consequently, they cannot be used to facilitate the development of new data systems to meet the real-world data management challenges of microservices. 


\section{The Online Marketplace Benchmark}
\label{sec:benchmark}








\subsection{Design Goals}


To bridge the aforementioned gap, it is key to pick an application scenario that covers all the aforementioned challenges. 
We find online marketplaces meet the following design goals.

\noindent\textbf{Industry strength.}
Online marketplaces present a substantial user base over the world~\cite{statista}. It is widely reported that online marketplace platforms employ microservice architectures to achieve benefits such as loose-coupling, fault tolerance, workload isolation, higher data availability, scalability, and independent schema evolution~\cite{b2w_eng,olist,vtex,broadleaf}.


\noindent\textbf{Low reproducibility barrier.} 
In the same spirit of the design of TPC-C \cite{tpcc}, online marketplaces fall into an application domain that can be easily understood by non-domain experts. This, in turn, reduces the learning curve and encourages benchmark adoption.

\noindent\textbf{Generalizability.} Many other popular application domains can be mapped to an online marketplace. These include multi-tenant e-commerce platforms, airline retailing~\cite{iata} (e.g., hotels, travel agencies, and taxi and insurance companies as sellers), social media marketplaces, 
delivery service (e.g., connecting individuals to a myriad of businesses, like restaurants), digital investment (e.g., funds from different banks as products), and homestay platforms.
It is noteworthy that many of these applications report data management challenges that we address in this work: event ordering and data invariants~\cite{nubank-arc, netflix_eda, uber_proxy,uber_money,airbnb_double, wix_pitfalls, wix}
transactional consistency across microservices~\cite{uber_microservice,airbnb_integrity,uber_payment,nubank,jet},
consistent data querying
~\cite{b2w_query,ifood,podium}
data replication correctness~\cite{synapse,netflix_eda,netflix_delta}.


In sum, an online marketplace is a suitable use case to benchmark data management in microservices.

\subsection{Application Scenario} 
\label{sub:scenario}

\noindent\textbf{Cart Management.} Customers shop in \OM by navigating and selecting products from a catalog. A customer session is linked to a \textit{cart}, with operations involving adding, removing, and updating items (e.g., increasing the quantity). A customer can only have one active cart at a time. 
When requesting a checkout, the customer must include a payment option and a shipment address, which are assembled together with the cart items and submitted for processing. 
On the other hand, customers may also abandon their carts before submitting the checkout request. 

\noindent\textbf{Catalog Management.} 
Each product is offered by a particular seller. Sellers are responsible for managing their products and associated stock information. They may replenish products and adjust their prices. 

\noindent\textbf{Customer Checkout.} Checkout requests follow a chain of actions: stock confirmation, order placement,
payment processing, and shipment of items. Upon submission of an order, if one or more items do not have sufficient stock, the checkout will still proceed with the available items. 

\noindent\textbf{Payment Processing.} Upon stock confirmation, the payment details provided by the customer are used to make a payment request to an external payment service provider (PSP)~\cite{psp}. A payment can fail for two reasons: rejection of the transaction by the provider or impossibility of contacting the payment provider.

\noindent\textbf{Order Shipment.} Upon approval of payment, the shipment process starts. For each seller present in an order, a shipment request is created. A shipment request includes the items present in the order. Each item is reflected as a package that should be delivered to the customer at some point.

\noindent\textbf{Package Delivery.} Whenever a package is delivered, both the corresponding customer and seller are notified. When all packages of an order are delivered, the order is considered completed.

\subsection{Microservices}
\label{sec:microservices}

\begin{figure*}
\centering
 \subfloat[\centering Partial Data model]{
  \includesvg[width=0.55\textwidth]{figures/draft_partial_model_0}
  }
  \hfill
\subfloat[\centering Events]{
  \includesvg[width=0.43\textwidth]{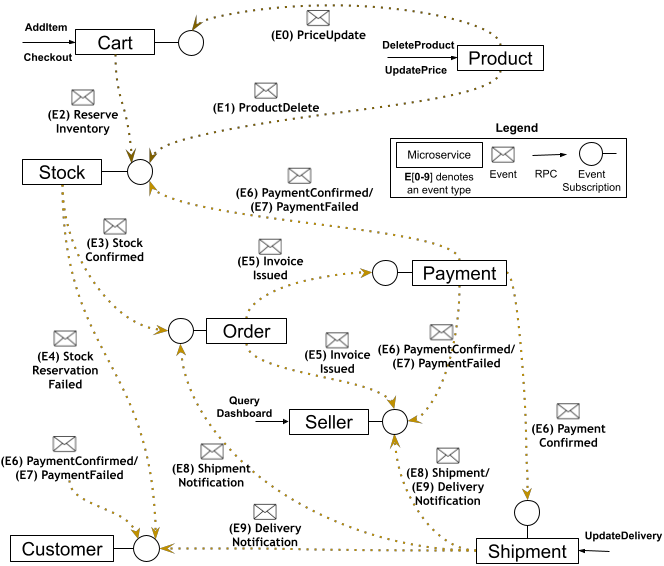}
  }
  \vspace{-2ex}
\caption{Online Marketplace Microservices}
\label{fig:marketplace}
\vspace{-3ex}
\end{figure*}

Traditionally, a microservice-based application is often made up of independent services, each deployed in a container on an OS-level virtualized platform such as Docker~\cite{vldb2021}. However, the emergence of programming frameworks for designing distributed applications, such as Orleans~\cite{bykov2010orleans} and Statefun~\cite{statefun}, provides programming models that allow microservices to be implemented using higher-level abstractions, such as actors and functions, respectively. 

To allow this benchmark to support the gamut of microservice implementations, we withdraw ourselves from defining an architectural blueprint, but rather, we focus on describing the expected independent components that must compose the benchmark application. In this sense, in the following paragraphs, we take advantage of Figure~\ref{fig:marketplace} to clarify the interactions, APIs, and the data model prescribed for each microservice. It is worth noting that we use the term "events" as a communication abstraction for microservices, but these can also be framed as any message payload asynchronously delivered to a microservice.
Besides, the event identifiers in Figure~\ref{fig:marketplace}(b) do not imply a particular order in which the microservices exchange events. 
Furthermore, while we specify the microservices' state and the queries using the relational model, they can also be specified through other data and query models as long as the same functionalities can be achieved.

\noindent\textbf{Cart.} The \textit{Cart} microservice allows the customer to manage the products that ought to compose a checkout order. It does so through APIs for managing cart items and submitting cart checkouts.

To manage a customer's cart, two relations are used: \texttt{carts} to track the status of a customer session, and \texttt{$cart\_items$} to store the items that customers want to buy. Moreover, the \texttt{$replica\_products$} relation represents a partial replica of the \textit{product} table belonging to the \texttt{Product} microservice. The table is only updated through \texttt{ProductUpdate} events received from the \texttt{Product} microservice (represented by a dotted blue arrow in Figure~\ref{fig:marketplace}(a)). The replication semantics are discussed in Section \ref{subsec:replication}.

When a checkout request is received, \texttt{Cart} ensures the correctness of the customer’s cart by matching the cart items to the replicated product information and looking for products that had their price updated. In this case, the price is updated and the difference (if positive) between the new and the old price becomes a discount in the cart item. The customer’s cart items are then assembled into a \texttt{\textit{ReserveInventory}} event and published for asynchronous processing. Once the event is published, the cart is sealed, and the customer is able to initiate a new cart session. 

\noindent\textbf{Product.} The \textit{Product} microservice manages the catalog of products. It performs operations over a single relation: \texttt{product}. For every update operation triggered via the following API, a corresponding event is generated for downstream processing: E0 for UpdatePrice and E1 for DeleteProduct.

\noindent\textit{DeleteProduct}. Marks a product as unavailable to customers. \newline
\noindent\textit{UpdatePrice}. Update the price of a product.

\noindent\textbf{Stock.} The \textit{Stock} microservice manages the inventory through a single relation: \texttt{$stock\_item$}. 
To update the inventory, the \textit{Stock} microservice processes four events: \texttt{ReserveInventory}, \texttt{PaymentConfirmed}, \texttt{PaymentFailed}, and \texttt{ProductDelete}. Every inbound event leads to updates in one or more \texttt{$stock\_item$}' tuples (depending on the number of items present in the checkout). 


On processing a \texttt{ProductDelete} event, the \textit{Stock} microservice marks the respective \texttt{$stock\_item$} as unavailable, meaning that future checkouts (triggered by \texttt{ReserveInventory} event) on this item cannot be carried out (represented by a red connector in Figure~\ref{fig:marketplace}(a)).

Processing the \texttt{ReserveInventory} event leads to marking the stocks of some items (subject to availability) presented in the customer's cart as reserved. The result of processing \texttt{ReserveInventory} might lead to the generation of two events: (i) \texttt{StockConfirmed} if at least one product has been reserved, and; (ii) \texttt{ReserveStockFailed} if no product requested by the customer is available.
In both cases, the respective items are assembled together in the respective event for downstream processing. Lastly, \texttt{PaymentConfirmed} confirms the reservation and \texttt{PaymentFailed} would lead to withdrawing the corresponding stock reservations.

Although at first sight, the differences between Product and Stock may seem blurry, it is important to highlight that they provide distinct functionalities in the application. While the \texttt{Product} is responsible for the correctness of the catalog data, including but not limited to the characteristics of a product like name, category, seller, and price, the \texttt{Stock} microservice ensures the integrity of the inventory. It is important to address that such design reflects real-world deployments~\cite{pstore}. 


\noindent\textbf{Order.} The \textit{Order} microservice manages customer orders' data. It does so by maintaining four relations: \textit{order}, $customer\_orders$, \texttt{$order\_items$}, and \texttt{$order\_history$}. The \textit{order} relation contains general data about an order, including but not limited to customer, amounts, etc. \texttt{$customer\_order$} relation tracks the number of orders requested by each customer, which is used to form the invoice number. Order items relate to the products requested in checkout and \texttt{$order\_history$} tracks updates to an order  (invoiced, paid, shipped, completed). Order's relations are updated upon processing the following events: \texttt{StockConfirmed}, \texttt{ShipmentNotification}, \texttt{PaymentConfirmed}, \texttt{PaymentFailed}.

Upon receiving a \texttt{StockConfirmed} event,  the amount to charge the customer is calculated (including freight and discounts) and an invoice number is generated, resulting in a \texttt{InvoiceIssued} event. During this processing, all four relations have tuples created (in case it is the first customer order, otherwise \texttt{$customer\_order$} tuple is updated). Afterwards, there are two moments where a customer’s order status is updated: after a payment and after shipment updates. Both trigger writes to order and \texttt{$order\_history$} relations. In this case, the order of event processing here plays a role on maintaining the notion of time progress for individual orders.

\noindent\textbf{Payment.} The \textit{Payment} microservice is responsible for managing payment data. Two relations are used to track an orders' payment data: $payment$ and $card\_payment$. Payment processing starts through processing an \texttt{InvoiceIssued} event. Every credit applied to the order, namely, discounts and the payment option chosen by the customer at the time of the checkout (credit/debit card or a bank slip) are stored in $payment$ relations and in case of a card payment, a $card\_payment$ tuple is also inserted.

As part of the payment processing, it may be necessary to coordinate with an external payment provider to ensure the provided payment method is valid. In this sense, \textit{Payment} microservice relies on an idempotent API offered by the external payment provider. In case the payment microservice fails right after confirming a payment, subsequent attempts (after recovering from failure) to charge the customer does not incur in duplicate payments.

Processing an invoice can lead to two outbound events: \texttt{PaymentConfirmed} and \texttt{PaymentFailed}. The \texttt{PaymentConfirmed} event contains all items present in the paid invoice whereas the \texttt{PaymentFailed} is a simple payload containing the invoice number so the \textit{Order} microservice can update its state accordingly.

\noindent\textbf{Shipment.} The \textit{Shipment} microservice manages the lifecycle of a shipment processing, including shipment provision and delivery of goods. Two relations are maintained: \textit{shipment} and \textit{package}. 

A shipment process starts by processing a \texttt{PaymentConfirmed} event, creating a delivery request for each order item and confirming all goods have been marked for shipment by producing a \texttt{ShipmentNotification} with status ‘approved’.

At a later moment, 
through the \textit{UpdateShipment} API,
a set of shipments and associated packages tuples are updated. Each package delivered leads to the generation of a respective \texttt{DeliveryNotification} and, in case all items that form a shipment are delivered, a \texttt{ShipmentNotification} with status ‘concluded’ is also emitted.


\noindent\textbf{Customer}. The \textit{Customer} microservice manages customer data, including their home address, contact information, and credit card. Besides, statistics about the customer are updated through processing the following events: \texttt{PaymentConfirmed}, \texttt{PaymentFailed}, \texttt{ReserveStockFailed}, and \texttt{DeliveryNotification}.

\noindent\textbf{Seller}. The \textit{Seller} microservice manages seller-based marketplace data. It does so by processing events and transforming them into seller-centric data. Three relations are found in \textit{Seller’s} schema: \textit{seller}, $order\_entry$, and $order\_entry\_details$.

The relation \texttt{$order\_entry$} represents an order item but from the perspective of a seller. In this sense, amounts related to the item (i.e., discount, freight, total) are calculated. Every order (derived from the \texttt{InvoiceIssued} payload) is transformed into $N$ \texttt{$order\_entry$} tuples, where N is the number of items corresponding to a seller. 
The relation $order\_entry$ is updated through processing the following events: \texttt{PaymentConfirmed}, \texttt{PaymentFailed}, \texttt{ShipmentNotification}, and \texttt{DeliveryNotification}.


\subsection{Workload}

In this section, we describe the \OM workload, namely, the business transactions and continuous queries the application must cope with.

\subsubsection{Business Transactions}
\label{subsec:business_transactions}

To realize the application scenario, we describe four business transactions that reflect different complexities in terms of the number of microservices involved and the number of events processed. In Section~\ref{sec:criteria}, we discuss how transactions possess different properties.

\noindent\textbf{Customer Checkout.} It starts in \textit{Cart} microservice through processing a \texttt{Checkout} request. It involves \textit{Cart}, \textit{Stock}, \textit{Order}, \textit{Payment}, \textit{Shipment}, \textit{Seller}, and \textit{Customer} microservices. A success path involves the following events: E2 - E3 - E5 - E6 - E8. There are two error cases: (i) when a payment fails, then from E5 we have E7 as the last; and when no single item from the cart is available in stock: E2 - E4. Although not shown in Figure~\ref{fig:marketplace} due to space constraints, a success path also involves E6 being processed by \textit{Order}.  

\noindent\textbf{Price Update.} To enable the partial replication of products in the \textit{Cart}, upon processing a \texttt{UpdatePrice} request and updating its private state accordingly, the \textit{Product} microservice generates E0 and sends it to the \textit{Cart} microservice. By processing E0, the \textit{Cart} microservice applies the new price to its corresponding replica.

\noindent\textbf{Product Delete.} 
To simulate making a product unavailable to customers, we pick a seller and a corresponding product (both from a distribution) and we set the product as disabled. To maintain the total number of products, thus avoiding anomalies in the distribution, we replace the deleted product with another one. This operation is done by processing a \texttt{DeleteProduct} request in \textit{Product} microservice. Upon updating its private state accordingly, \textit{Product} generates E1 and sends it to \textit{Stock}.


\noindent\textbf{Update Delivery.} 
To simulate the delivery of goods, we pick the first 10 sellers with uncompleted orders (i.e., at least one package has not been delivered yet) in chronological order, and we set their respective oldest order's packages as delivered. Thus, packages are progressively delivered as more update delivery transactions are submitted to the system. For maintenance of statistics about customers and orders, the following events are generated with the transaction:
\texttt{E9} is generated by \textit{Shipment} for every package delivered and sent to the Seller and Customer microservices; and \texttt{E8} is generated by \textit{Shipment} when all packages of a shipment are delivered and sent to \textit{Order}.

\lstset{style=myJava}


\subsubsection{Continuous Queries}
\label{subsec:queries}

Continuous queries over event streams constitute an emergent trend in microservice applications. As event payloads produced by different microservices often contain state information~\cite{eventdrivenfowler}, it is possible to (indirectly) access data from multiple microservices without breaking their encapsulation. In other words, continuous queries can be built based on events without resorting to (synchronously) pulling data from each required microservice. 

In this section, we describe three different types of continuous queries covering important concerns in \textit{Online Marketplace}. To specify the queries, we use the syntax of Materialize~\cite{materialize}, a streaming database.
Due to space constraints, we explain one continuous query and the others can be found in our extended version~\cite{extended}.


%




\noindent\textbf{Query \#1: Seller Dashboard.}
The business scenario encountered in a marketplace requires sellers to have an end-to-end overview of the operation in real-time to identify trends, such as the popularity of products, and to support decision-making, such as when to increase product prices.

\noindent\textbf{Query Description:} In this query, we want to determine the total financial amount of ongoing orders by seller. Assuming there is a stream (Listing 1) that filters out concluded and failed orders, one could implement this continuous query as shown in Listing 2.

\begin{lstlisting}[language=SQL, caption=Ongoing order entries base query]
CREATE MATERIALIZED VIEW order_entries
SELECT order_id, seller_id, product_id, ...
FROM InvoiceIssued as inv
LEFT JOIN ShipmentNotification as ship ON ship.order_id = inv.order_id
LEFT JOIN PaymentFailed as pay ON pay.order_id = inv.order_id
WHERE ship.status != 'concluded' AND pay.order_id IS NULL
\end{lstlisting}

\begin{lstlisting}[language=SQL, caption=Ongoing orders aggregation per seller]
SELECT seller_id, COUNT(DISTINCT order_id) as count_orders, COUNT(DISTINCT seller_id, product_id) as count_items, SUM(total_amount) as total_amount, SUM(freight_value) as total_freight, SUM(total_items - total_amount) as total_incentive, SUM(total_invoice) as total_invoice, SUM(total_items) as total_items
FROM order_entries
GROUP BY seller_id
\end{lstlisting}

In addition, the seller dashboard output must also discriminate the records that compose the aggregated values computed in the above query. In this sense, it should also present to the user the following query result:

\begin{lstlisting}[language=SQL, caption=Ongoing orders discriminated]
SELECT * FROM order_entries 
WHERE seller_id == <sellerId>
\end{lstlisting}

\noindent\textbf{Query \#2: Cart Abandonment.}  A popular use case arising from a series of customer interactions is cart abandonment~\cite{abandoned}. A cart is considered abandoned in two cases: (i) prior to checkout submission, and (ii) upon a failed payment processing, if no customer checkout re-submission is identified.




\noindent\textbf{Query Description:} 
In this query, we want to find the cart checkouts that have either failed via stock reservation or payment attempt and have not been involved in a new checkout attempt within the next 10 minutes after the failure. Upon detecting an abandoned cart, a \texttt{CartAbandoned} event is generated for both \textit{Cart} and \textit{Customer} microservices. The query specification can be found in our extended version~\cite{extended}. 


\noindent\textbf{Query \#3: Low Stock Warning.}
To assist sellers, marketplaces usually monitor products' stock proactively to notify sellers about low inventory and ultimately refrain customers from experiencing unavailability of products.

\noindent\textbf{Query Description:} 
In this query, we want to find the products (and their respective sellers) that are likely to face unavailability in case no replenishment is provided in the near future.
The search is based in the following criteria:
The average number of items requested per week in the last month, independently of the result of the reservation and payment, is higher than the present stock level.
The \textit{Low Stock Warning} continuous query specification can also be found in our extended version~\cite{extended}.

\section{Data Management Criteria}
\label{sec:criteria}

In this section, we discuss the criteria that an implementation of \OM\ should meet. The criteria reflect key principles and challenges of data management in microservices. Some criteria have multiple levels, allowing the benchmark users to choose the most suitable one that fits their specific requirements. Furthermore, our explicit criteria specifications facilitate conducting fair comparisons between different systems on the same basis. 



\subsection{Communication and Data Access}

Microservices are functionally partitioned applications~\cite{base}. Two issues arise with this model:

\noindent(i) Microservices interact with each other through asynchronous events to carry out functionalities across multiple microservices. For example, completing a cart requires a composition of functionalities across \textit{Cart}, \textit{Stock}, \textit{Order}, \textit{Payment}, and \textit{Shipment} through the events they exchange.



\noindent(ii) State management is divided into two categories: (a) Direct data access: Each microservice can directly access its own data (encapsulation principle); and (b) Indirect data access: Data within each microservice can only be accessed externally through predefined interfaces (e.g., seller dashboard) or be notified about state changes through events (e.g., \textit{Cart} is notified about product updates). It should be noted that this criterion does not require data from different microservices to be stored in different databases.




\subsection{All-or-nothing Atomicity} 

The business transactions must comply with all-or-nothing atomicity semantics. This criterion is crucial to guard against crashes or performance degradation leading to failures in the middle of a business transaction. 







\subsection{Caching or Replication} 
\label{subsec:replication}
Section \ref{sub:scenario} describes the case of \textit{Cart} subscribing to product updates with the goal of ensuring that checkouts do not contain outdated product prices. As the strategy of implementation varies system by system, either cache or replication can fulfill this requirement. In this sense, we prescribe three possible correctness semantics:

(i) Eventual. Updates (price update or product delete) are processed independently, disregarding the order that they are generated at the source, i.e. \textit{Product}. 

(ii) Causality at the object level. Updates on the same product are processed sequentially in accordance with the order in which they were performed at the source. 

(iii) Causality across multiple objects. All updates made by the same seller must be applied to the cart in the same order they were applied at the source, achieving read-your-write consistency. 

\subsection{State Safety Properties}

\subsubsection{Inter-microservices Properties}. Refers to a property that cut across microservices.

\noindent\textbf{S\#1: Cross-microservice referential integrity.} \textit{Stock} always references an existing product in \textit{Product}. We take inspiration from the CAP theorem~\cite{cap} to define two levels of consistency:

(i) Available System. A deleted product will eventually not be allowed to be reserved anymore. The transaction is considered committed after \textit{Product} responds to the \texttt{Delete Product} request. 

At this level, if there is a network error or failure of the microservices, the delete message (EY) may not reach \textit{Stock}. That requires the inconsistency to be detected and resolved at a later point.


(ii) Consistent System. \texttt{Delete Product} request is committed only when both Product and Stock have committed, therefore Product and Stock are always consistent with each other. At this level, a deleted product will not be available for future checkout attempts. 




\noindent\textbf{S\#2: No duplicate checkouts.} 
The cart of a customer session must not be checked out more than once. This can be safeguarded by having mutual exclusion in each cart. Another way to ensure this property is by assigning a customer session ID to each new customer session. The ID is included in the payload of each checkout request submitted to \textit{Order}. Upon receiving it, \textit{Order} ensures the same cart checkout does not lead to duplicate order processing.

\subsubsection{Intra-microservice Properties}
These include properties that can be enforced by relying on only local states. They can be achieved through appropriate isolation levels. 

\noindent\textbf{S\#3: No overselling of items.}
In \textit{Stock} microservice, both available and reserved quantities cannot fall below zero for every item. In addition, the reserved quantity must never be higher than the available quantity.


\noindent\textbf{S\#4: Maintenance of customer statistics.}
The system tracks successful and failed payments and deliveries through increments of numbers. An appropriate isolation level is required to ensure that the increment in \textit{Customer} is not missed.

\noindent\textbf{S\#5: Linearized product updates.} Updates on each individual product's price or version in the \textit{Product} state must be linearized.

\noindent\textbf{S\#6: Concurrent Update Delivery transactions must execute in isolation.}
Since several events can be generated during a \textit{Update Delivery} transaction, it is essential that concurrent \textit{Update Delivery} transactions do not operate on the records of the same shipments and packages. The purpose is to avoid emitting duplicated \texttt{ShipmentNotification} and \texttt{DeliveryNotification} events.


\noindent\textbf{S\#7: Stock operations triggered by events must execute in isolation.}
Processing events in \textit{Stock} requires either serializable isolation or exclusive locks on the items contained in the event payload.

\vspace{-3ex}
\subsection{Event Processing Properties}

\noindent\textbf{Event Order.} Event processing order impacts \textit{Order} and \textit{Seller} dashboard. A microservice data platform can provide two levels of event ordering guarantee.

\noindent Unordered: \texttt{InvoiceIssued}, \texttt{PaymentConfirmed} and \texttt{ShipmentNotification} events are processed arbitrarily, possibly leading to violating the natural order processing workflow. \newline
Causally Ordered: \texttt{InvoiceIssued} must precede payment events. The \texttt{PaymentConfirmed} event must always precede shipment events. Similarly, a \texttt{ShipmentNotification} with the status ‘approved’ and \texttt{DeliveryNotification} must always precede the corresponding \texttt{ShipmentNotification} with the status ‘concluded’.

\noindent\textbf{Consistent Snapshot for Continuous Queries.} For the two concurrent queries of the seller dashboard to be consistent with each other, their results should reflect the same snapshot of the application state.




\noindent\textbf{Event Delivery.} The data platform can provide three levels of event delivery guarantees, namely at-most-once, at-least-once, and exactly-once delivery. The delivery guarantee has impacts on how to implement continuous queries and business transactions in \OM\ to achieve correctness. At-least-once and at-most-once delivery can make queries inaccurate. For example, while at-most-once delivery provides a lower bound on the profits in the \textit{Seller Dashboard} without replaying lost messages, at-least-once can provide skewed results without accounting for duplicate events.

For transactions, at-least-once delivery requires microservices to be idempotent to account for possible duplicate event processing. On the other hand, a timeout mechanism must abort transactions if at-most-once or exactly-once delivery is used to prevent stalled transactions from blocking the other transactions.

\subsection{Performance and Failure Isolation} 
\label{subsec:isolation}

The functional partitioning of the application allows microservices to operate independently, offering benefits from the isolation of resources assigned to each microservice~\cite{base,olep}. Implementations of \OM\ can achieve isolation in two tiers: 

(i) Each microservice's code/logic is executed on different computational resources, achieving performance and fault isolation at the application tier. This minimizes the impact of resource usage and failures between different microservices.

(ii) In the database tier, database operations do not interfere with each other in terms of performance and failure, achieved through isolated resource allocation to databases.




\subsection{Logging}

Audit logging is a critical concern in applications as it allows developers to track application events like user activities (e.g., checking out a cart) and state updates~\cite{datadog}. These events recorded during application execution serves as an audit trail, aiding developers in troubleshooting faults and verifying compliance with prescribed business rules. This is even more pressing in distributed systems, such as microservices. 
The substantial exchange of events and the complex interplay of independent components make it challenging to reason about errors and failures involving multiple asynchronous microservices. 
Thus, data systems must store audit events durably.
In \OM, there are two key events related to logging historical records of operations: \texttt{ShipmentNotification} and \texttt{PaymentFailed}.

\textbf{(a)} Upon \texttt{ShipmentNotification} with \texttt{status} 'completed' or \texttt{PaymentFailed}, \textit{Order} logs all records associated with such an order, in particular the relations \textit{order}, \texttt{$order\_items$}, and \texttt{$order\_history$}.

\textbf{(b)} Upon \texttt{ShipmentNotification} with \texttt{status} 'completed', \textit{Seller} logs all records associated with such shipment, in particular the relations \texttt{$order\_entry$} and \texttt{$order\_entry\_details$}.

\textbf{(c)} As part of the emission of a \texttt{ShipmentNotification} with \texttt{status} 'completed', \textit{Shipment} logs all records associated with such shipment, in particular those in the relations \textit{shipment} and \textit{package}.

\textbf{(d)} As part of payment processing, \textit{Payment} logs payment records independently of the outcome (success or failure), particularly the relations \textit{payments} and \textit{$payment\_cards$}.


\section{Data and Workload Composition}
\label{sec:workload}




The workload submitted to \textit{Online Marketplace} can be adapted to fit particular needs. In the following, we present the different configuration parameters that can de defined for experiments.

\noindent\textbf{Data Population.} The state of some of the microservices require initialization prior to workload submission. The following procedure is expected, where $X$, $Y$, and $Z$ are configuration parameters:\newline
(i) $X$ number of customers are inserted into \textit{Customer} microservice; \newline
\noindent (ii) $Y$ number of products are inserted into \textit{Product} microservice; \newline
\noindent (iii) $Y$ number of stock items are inserted into \textit{Stock} microservice. Each $stock\_item$ tuple must refer to an existing product (through $seller\_id$ and $product\_id$ columns) in \textit{Product} microservice; \newline
\noindent (iv) $Z$ number of products per seller leads to ($Y$/$Z$) seller tuples inserted into \textit{Seller} microservice. In case of positive remainder, the last seller must own W products, where $W$ < $Z$. 

On the one hand, the number of customers should be large enough to accommodate the maximum amount of concurrent transactions running the system at a given time. For instance, considering a throughput $T$, $X$ must be higher than $T$ to guarantee there is always a customer available. On the other hand, the number is limited by the amount of memory available.

Furthermore, the number of products and sellers should be large enough not to create many conflicts when running. 
Considering $prob\_s$ and $prob\_p$ as the probabilities of picking a seller and of picking a seller's product, respectively, to have less than $\delta$ conflicts, one should pick distributions that lead to $T$ * $prob\_s$ * $prob\_p$ < $\delta$.

To aid data population, the dataset can be generated based on a given \textit{size factor} following an uniform distribution. A size factor $S$ leads to 10K customers ($X$), products and also associated stock items ($Y$).
Benchmark users can populate records in other microservices to simulate preexisting data and introduce a degree of overhead in state accesses during the workload execution. A size factor $S$ leads the \textit{order} table to be initialized with $S$*100K tuples uniformly distributed among customers, having the number of \textit{order item}s randomly picked (1-10). In consequence, \textit{payment}, and \textit{shipment} follow the same size given their tuples refer to an existing order. 

\color{blue}
\color{black}
\noindent\textbf{Workload Distributions.} Distribution in the workload is centered on sellers and their products. For every operation involving a product (add cart item, product price update, and product delete), one has to pick first a seller and then proceed to pick a corresponding item from the seller's product keyset. In this way, Uniform and Zipfian distributions can be used interchangeably in two cases: \newline
- Seller selection: for every product selection, one has to first pick a seller based on a defined distribution. \newline
- Product selection: After a seller is previously selected, one picks a seller's product based on a defined distribution.

\noindent\textbf{Metrics.} We collect two metrics in this benchmark. Throughput is the number of transactions processed per second, which includes both business transactions that have been successfully completed and continuous queries that have successfully returned a result. 
End-to-end latency is measured from the moment a client sends a transaction request until the result is received.

\noindent \textbf{Transaction ratios} define the probabilities of clients submitting various types of business transactions and continuous queries to the system. We envision it can be set to reflect some potential scenarios:

\textbf{A. Order-heavy scenarios} reflect e-commerce scenarios and high-sales periods, the workload is dominated by order processing (i.e., checkouts and deliveries), having a small percentage of abandoned carts and payments declined. This workload includes both uniform and non-uniform seller distributions, which models for example, some sellers' products being much more popular than others, leading to orders being canceled due to unavailability. 
The parameters are inspired by Olist workload~\cite{olist_dataset} and specified  as product delete with 2\%, price update with 3\%, seller dashboard with 5\%, customer checkout with 30\%, and update delivery with 60\%.

\textbf{B. Update-heavy scenarios} reflect frequent updates with low-latency requirements, as found in banking, financial trading, and airline applications. 
This workload models a substantial degree of product updates with non-uniform seller and product distributions. 
Varying the zipfian constants allows for controlling the number of accesses to Cart and Stock, similar to some extended versions of Smallbank~\cite{reactdb,snapper}.
One can set the zipfian constants ranging from 0.2 to 1.2, and transaction ratio as product delete and price update with 30\% each, seller dashboard with 2.5\%, customer checkout with 12.5\%, and update delivery with 25\%.

\textbf{C. Hybrid transactional-analytical scenarios} model a substantial number of users continuously querying their data (i.e., the seller dashboard) while the system must also cope with moderate transactional requests. The goal is to prevent the use of simplified caching mechanisms. Thus, seller distribution is uniform while products may have skewed access. Transaction ratios are inspired by HTAP benchmarks~\cite{CH,HyBench} and defined as seller dashboard with 48\%, customer checkout with 16\%, update delivery with 32\%, product delete and update price with 2\% each.

\section{Benchmark Driver}


\subsection{Driver Functionalities} 
\label{subsec:driver_func}


To manage the life-cycle of an experiment composed by: (i) data generation, (ii) data ingestion, (iii) data check, (iv) workload submission, (v) collection of results, (vi) report generation, and (vii) cleaning of states, we develop a benchmark driver in .NET~\cite{extended}.

A user specifies the data population, workload composition, ratio of transactions, and distributions through a configuration file. A user usually starts with generating data. Sellers, customers, products, and stock items are generated using Bogus~\cite{bogus}, a popular library for generating synthetic but realistic data, including email, full name, address, and social security number. For products, Bogus generates matching product names and descriptions, including prices and locations.
Generated data can be either stored durably or kept in main memory via DuckDB~\cite{duckdb}. Users can load the data from DuckDB and move on to data ingestion, populating the microservices with initial data.
Further details of the driver software, including workload generation, can be found in the extended version of this paper~\cite{extended}.

\subsection{Driver Implementation Challenges}
\label{subsec:driver_challenges}

The \OM\ benchmark presents particular characteristics that necessitate caution on submitting transactions.

\noindent\textbf{Simulating Customer Sessions.} We start with the \textit{Cart} microservice, which can be exemplified as a stateful operator. In the context of a customer session, a cart state evolves by having sequential operations (i.e., add cart item) up to a point where the state is sealed (through a checkout request), then returning to the initial state. To this end, the workload submission must account for each customer session individually. In case there is an active customer session for customer \texttt{X}, \textit{there should be no concurrent driver's thread simulating the same customer}. The reason is that concurrent operations on  \texttt{X}'s cart state can interleave, making it difficult to maintain the workload distribution.


To meet this requirement, a thread-safe queue for idle customers is maintained. Whenever a thread initiates a checkout transaction, it picks a customer from the idle queue to represent. After the checkout request is submitted, the thread pushes the customer to the queue, making it available again for other threads. The time a customer lies in the idle queue is driven by the concurrency level supported by the target platform. The faster checkouts are processed, the less time customers spend in the idle queue.

\noindent\textbf{Managing Coherent Product Versions.} 
As explained in Section~\ref{subsec:business_transactions}, products could be deleted and replaced by new ones. In other words, new product versions are being generated online continuously. It is crucial to generate transactions referring to coherent product versions.
For example, to generate a price update transaction, we need to make sure to refer to the current product version in \textit{Product}'s state. As another example, on adding an item to a cart, we need to make sure cart items refer either to the current product version in \textit{Product} or the version that preceded the current one at the time of this operation. The goal is to simulate the latest product version "seen" by a customer.

To obtain a coherent product version while constructing transactions, we have to know the application state. However, the benchmark driver should not query microservice states during transaction submission for two reasons:
(i) Workload generation should be independent of the actual data platforms being used;
(ii) Querying the state of microservices would introduce additional load that is not prescribed in the benchmark.

In order to generate correct transactions while not querying the microservices' states, the driver manages internally a consistent mirror of the \textit{Product} microservice state to guarantee access to coherent product versions. 
The driver linearizes the submission of concurrent update requests to the same product; the compare-and-swap mechanism is used to decrease synchronization costs. 

Whereas the picking of product versions for building a cart item runs concurrently with product updates so that customer and seller threads do not block each other.

\noindent\textbf{Matching Transaction Requests to Asynchronous Results.} It is common that platforms for building microservice applications provide results asynchronously (Section ~\ref{subsec:driver_func}).
In such an asynchronous system, the driver must track each submitted transaction and match it with the corresponding asynchronous transaction result to compute the metrics accurately. To this end, each transaction request is assigned a timestamp and a unique ID. This ID is later used to match the request to a corresponding transaction result that is eventually received. To avoid threads blocking each other, ID generation is decoupled from transaction submission. In conclusion, a stateful driver is necessary to provide correct transaction input.
\vspace{-2ex}
\section{Case Studies}
\label{sec:platforms}

To demonstrate the applicability of \OM\ and show the design implications and effects on different systems, we implement and experiment five versions of \OM\ in three competing platforms, Orleans, Statefun, and a custom solution. Orleans and Statefun are designed to develop event-driven services with state management functionalities. The custom solution is based on a combination of multiple systems that reflects an usual architecture seek in practice~\cite{vldb2021,olep}, which will also provide an insight into the complexity faced by the developers. From the implementations and experimental study, we derive lessons learned (referenced by \textbf{L\#}) and design decisions (\textbf{D\#} in $\S$~\ref{subsubsec:design}) to foment new systems' design and improve both state-of-the-art platforms, highlighting the usefulness of \OM.



\subsection{Implementations}

\subsubsection{Microsoft Orleans}

Orleans is a framework for building distributed stateful applications. In Orleans, applications are composed of concurrent virtual actors~\cite{bykov2010orleans}, each encapsulating a private state, that are allocated transparently across machines~\cite{orleans_silo}. Developers are offered APIs to log actor state~\cite{orleans_persistence}, which allows actors to recover their state upon crashes or server migrations. In our implementation, we use the default Orleans actors' memory storage API to store the microservices data.



Due to the single-threaded actor abstraction, developers are encouraged to decompose application functionalities into distinct actors to avoid a few specific virtual actors becoming the bottleneck. Inspired by the guidelines of Wang et al.~\cite{iot}, our design aims to maximize parallelism and minimize transaction latency by assigning \OM\ functionalities to different actors. Due to space constraints, further details about the components' design is found in our extended version~\cite{extended}. We implement \OM\ on Orleans 7.2.1, using both the default non-transactional Orleans API and the transactional API, referred to Orleans Transactions (TX) in the rest of the paper. To allow Orleans actors to be reachable from the driver, we deploy an HTTP server on top of the Orleans silo. The server is responsible for parsing and forwarding incoming requests to the appropriate actors, reporting back the transaction results via an HTTP interface.






\subsubsection{Flink Statefun}

Statefun is a platform built on top of Flink for running distributed applications based on the concept of stateful functions~\cite{statefun}. Flink processes manage the state, handle incoming requests, and invoke the appropriate functions~\cite{statefun_arc}. Each function encapsulates its own logical state and reacts to incoming messages asynchronously. Each incoming message is processed sequentially, in a way similar to Orleans actors. As a result, applications can be designed by composing functions through messages. Similarly to Orleans, we use the Statefun memory storage API to store the microservices data.

We implement \OM\ on Statefun 3.3, and, given the resemblance of both programming models, we opted to model stateful functions with the same design as Orleans. To match the architectural design found in web services, we deploy Statefun with an HTTP ingress. Once the driver submits the transaction input, the ingress acknowledges the reception of the request and dispatches it to the appropriate function. We also deploy an HTTP egress to allow for the collection of transaction results that are eventually completed. The function that terminates a transaction sends a completion message to the egress. The egress operator stores the result and makes it available to clients. This design follows the Statefun documentation~\cite{shopping-cart}. To execute Stateful Functions runtime, we follow the recommended deployment mode using docker images and the most performant execution style (co-located functions)~\cite{statefun_colocated}.


\subsubsection{Implementation Insights}

Upon implementing the \OM\ features in each platform, we encountered some limitations. We explain in the following how we mitigated them and the criteria that the target platforms can meet. Note that we only consider features that are natively supported by the framework, and use as few external systems as possible.

\noindent\textbf{Data replication.} Given that both platforms do not provide indexing for the actor or function states, we cannot efficiently query which carts contain a particular product. 
Although we could carry out this search iterating over all carts of the keyspace, we found this incurs a heavy cost operation, significantly impacting the benchmark analysis. Therefore, we opted not to implement this replication feature in the platforms. 

\noindent\textbf{Continuous queries.} We do not use external stream processing systems in order to benchmark only the target platforms rather than the integration of multiple systems. Therefore, we opted to only implement the query dashboard, but not the continuous queries of cart abandonment and low stock warning.
The reason is that the \texttt{order\_entries} view can be maintained through conventional 
data structures and updated through application events, while the others require more complicated stream processing operators such as windowed aggregates and joins, which are not provided natively by the two platforms. \newline
\textbf{L\#1.} \textit{Apart from being complex and error-prone, ad-hoc replication and continuous query implementations incur an overhead on both Orleans and Statefun platforms.}






\noindent\textbf{Messaging delivery guarantees.} Orleans provides an at-most-once delivery guarantee by default. Although Orleans can be configured to send retries upon timeout, we opted to capture the timeout exception and report to the driver that the transaction has been completed with an error. The reason is that, by enabling retries, the message may arrive multiple times, potentially corrupting data if the application is not idempotent~\cite{base}. 

On the other hand, Statefun manages state storage and message delivery in an integrated manner, such that in case of a delivery error, Statefun transparently retries the delivery up to a timeout and, upon that, rewinds the application to a previously consistent checkpoint~\cite{statefun_checkpoint}. To make the performance results of the two platforms comparable, we disabled checkpointing in Statefun, and in case of a delivery error, we proceeded in the same way as in Orleans. \newline
\textbf{L\#2.} \textit{The lack of exactly-once processing guarantees in Orleans requires developers to make their operations idempotent, adding execution overhead and increasing implementation complexity.}

\noindent\textbf{Logging.} We do not use Orleans storage because we run into the problem of inconsistent state while logging the actor's state~\cite{orleans_bug}. Instead, we use external PostgreSQL to log completed transactions on both platforms. The consistent mechanisms also make the results of the two platforms comparable.

\noindent\textbf{Resource Isolation.}
Both actors in Orleans and functions in Statefun share computational and data storage resources of the node they are allocated to. To isolate the computational resources of a specific microservice, it is necessary to configure a custom grain placement strategy in Orleans~\cite{grain_placement}. In Statefun, functions are allocated to Flink Task Managers transparently~\cite{flink_task_manager}. To make the results comparable, we use Orleans default placement strategy. \newline
\textbf{L\#3.} \textit{The lack of explicit interfaces for configuring function allocation in Statefun can jeopardize achieving performance and failure isolation for microservices.}





\subsubsection{A Composite Solution}

Our experience above matches recent findings that microservice practitioners must combine several heterogeneous systems to fulfill their data management requirements~\cite{vldb2021}. Inspired by these results, we use a popular stack of technologies that practitioners often use to design a full-featured \OM\ implementation.

To achieve all-or-nothing atomicity and concurrency control, we base our solution on Orleans Transactions. We offload consistent querying to PostgreSQL. With a combination of materialized view and refresh view triggered via actors 
, we achieve a consistent seller dashboard snapshot through a PostgreSQL transaction. 


We implement two replication semantics: (i) Eventual, enabled by Orleans Streams~\cite{orleans_streams}, an Orleans add-on that allows actors to subscribe and publish to streams dynamically; and (ii) Causality across multiple objects, enabled by a primary-secondary Redis deployment. To decouple writers from readers, we use the following scheme: \textit{Product} actors write to the primary node, and updates are asynchronously streamed to the secondary node, so \textit{Cart} actors can read from it. As Redis linearizes writes, updates are applied to the \textit{Cart} in order. 

\noindent\textbf{L\#4.} \textit{Achieving consistent replication and querying requires weaving together heterogeneous systems through non-native APIs.}

\vspace{-2ex}
\subsection{Experimental Study}
\label{sec:exp}

\subsubsection{Experimental Settings}
\label{subsec:settings}

\noindent\textbf{Deployment.} We set up our benchmarking environment on UCloud \cite{ucloud} based on u1-standard instances. A u1-standard contains an Intel Xeon Gold 6130 CPU@2.10 GHz, 32 CPUs, and 384 GB of memory~\cite{u1_standard}. UCloud instances run inside a Kubernetes cluster connected via 100Gbps Infiniband virtual network. To avoid resource competition, we allocate independent instances to different benchmarking components, namely, the benchmark driver, the target platform, the PostgreSQL database server, and the Redis instances. To remove cache effects, we restart Orleans and Statefun after each run. In experiments involving Redis and PostgreSQL, we also clean up the state after each run.
Besides, after data population, we ensure CPU usage returns to idle before initializing transaction submission.
We use PostgreSQL 14.5 running in the operating system (OS) Debian 12.1, whereas the driver and platform instances run in the OS Ubuntu 22.10. All the instances are located in the same region and availability zone. The criteria in $\S$\ref{subsec:isolation} are only applied to multi-node deployments. Since we focused our investigation on the performance and overhead of the platforms in multi-core settings, these are not applicable.

\noindent\textbf{Methodology.} All experiments are run in 6 epochs of 10 seconds each with the first 2 epochs as a warm-up period. We collect two metrics: throughput and end-to-end latency.
For each experiment, we maximized the resource allocated to the instances running the benchmark driver and PostgreSQL and we made sure resource usage was kept under 80\%, to avoid them becoming the bottleneck.

\noindent\textbf{Workload and Parameters.} 
We use the workload modeling \textbf{scenario A} ($\S$~\ref{sec:workload}).
We set the probability of payments accepted and checkout to 100\% to maximize the amount of transactions. Checkouts contain up to 10 items and quantities are selected randomly for each item (1-5). Each cart item has 5\% probability of being assigned a discount. The discount value is randomly picked, not surpassing 10\% of the item's price. 
We use both uniform and skewed distribution for picking sellers per each transaction. The number of sellers and products are picked reflecting \textit{size factor} 10,  so to not introduce substantial conflicts in uniform distribution.

\subsubsection{Experiment Results}



\begin{figure*}
\centering
  \begin{tikzpicture}[scale=0.62]
    \begin{axis}[
        width=8cm,height=3.5cm,
        scale only axis,
        xlabel={\textbf{(a)} Concurrency Level},
        ylabel={\# transactions},
        ylabel near ticks,
        xlabel near ticks,
        ymin=0, ymax=20000,
        xtick={1, 8, 16, 24, 32},
        ytick={0,5000,10000,15000,20000}, 
        legend style={legend pos=north west,font=\footnotesize},
        legend cell align={left},
        ymajorgrids=true,
        xmajorgrids=true,
        grid style=dashed
        ]
        \addplot
            coordinates {
            (1,598)
            (8,4773)
            (16,9554)
            (24,14327)
            (32,19110)
            };
        \addplot
            coordinates {
            (1,597)
            (8,4774)
            (16,9556)
            (24,14339)
            (32,18897)
            };
        \addplot
            coordinates {
            (1,590)
            (8,4723)
            (16,9441)
            (24,14159)
            (32,18496)
            };
        \addplot
            coordinates {
            (1,590)
            (8,4723)
            (16,9443)
            (24,14169)
            (32,18624)
            };
         \legend{UNIF-UNIF, UNIF-ZIP, ZIP-UNIF, ZIP-ZIP}
    \end{axis}
\end{tikzpicture}
  \begin{tikzpicture}[scale=0.62]
    \begin{axis}[
        width=8cm,height=3.5cm,
        scale only axis,
        xlabel={\textbf{(b)} Concurrency Level},
        ylabel={Throughput (tx/s)},
        ylabel near ticks,
        xlabel near ticks,
        ymin=0, ymax=9000,
        xtick={1, 10, 20, 30, 40, 50, 60},
        ytick={0,2000,4000,6000,8000},
        legend style={legend pos=north west,font=\footnotesize},
        legend cell align={left},
        ymajorgrids=true,
        xmajorgrids=true,
        grid style=dashed
        ]

        \addplot
            coordinates {
            (1,674)(8,3078)(16,4965)(24,6141)(32,6471)(40,6729)(48,7144)(56,7127)
            };
        \addplot
            coordinates {
            (1,591)(8,2688)(16,4354)(24,5446)(32,5751)(40,6000)(48,6264)(56,6300)
            };
        \addplot
            coordinates {
            (1,281)(4,613)(8,665)(12,520)(16,871)(24,724)(32,793)(48,718)(56,796)
            };

        \legend{Orleans, Orleans+Logging, Orleans TX}
    \end{axis}
\end{tikzpicture}
  \begin{tikzpicture}[scale=0.62]
    \begin{axis}[
        width=8cm,height=3.5cm,
        scale only axis,
        xlabel={\textbf{(c)} Concurrency Level},
        ylabel near ticks,
        xlabel near ticks,
        ymin=0, ymax=2500,
        ytick={0, 500, 1000, 1500, 2000, 2500},
        xtick={1,200,400,600,800,1000,1200,1400,1600},
        legend style={legend pos=south east,font=\small},
        legend cell align={left},
        ymajorgrids=true,
        xmajorgrids=true,
        grid style=dashed
        ]

        \addplot
            coordinates {
            (1,4)(8,39)(16,75)(24,103)(32,126)(48,190)(64,323)(128,748)(256,1674)(512,2302)(768,2156)(1024,2129)
            };
        \addplot
            coordinates {
            (1,4)(8,41)(16,79)(24,109)(32,170)(48,227)(64,331)(128,699)(256,1315)(512,1383)(768,1354)(1024,1473)
            };

        \legend{Statefun, Statefun+Logging}
    \end{axis}
\end{tikzpicture}
\vspace{-3ex}
\caption{Driver Scalability (a) and Concurrency Overhead (b \& c)}
\label{fig:concurrency}
\vspace{-3ex}
\end{figure*}
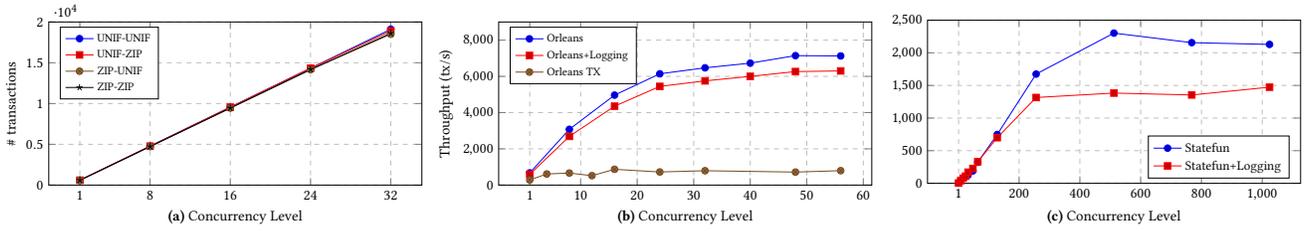

\noindent\textbf{Driver Scalability.}
To examine if the coordination mechanisms of the driver (Section~\ref{subsec:driver_challenges}) are scalable,
we conduct a scalability experiment. We use the parameters 
of \textbf{scenario A} ($\S$~\ref{sec:workload})
and fix the transaction latency to 100 ms, varying the number of concurrent workers submitting transactions. We observe in Figure~\ref{fig:concurrency}(a) that the number of completed transactions scales with the concurrency level, showing stable throughput across epochs. The overhead entailed by adding more worker threads does not create a bottleneck, independent of the distribution picked for sellers and products, reflecting the efficiency of the driver's coordination mechanisms.



\noindent\textbf{Effect of Concurrency Level.}
In this experiment, we refer to \emph{concurrency level} the maximum number of concurrent transactions running in the system at a given time. We measure the overhead incurred by different concurrency levels and identify the parameter that provides the optimal performance in each platform for further evaluation. In this section, we maximize the resources available (32 CPUs) and we set the workload skewness to be uniform for both sellers and products. 


As shown in Figure~\ref{fig:concurrency}(b), Orleans achieves maximum throughput 
with the concurrency level 48. With an even higher concurrency level, more messages would be accumulated on actors' input queue, particularly actors in the checkout critical path (e.g., with \textit{Order} and \textit{Payment} doubling their processing latency). 
The overhead introduced by logging is negligible at the start and remains constant as the concurrency level increases, achieving a maximum of 12\% overhead compared to non-logging. The lower throughput derives from the wait introduced by updates submitted to PostgreSQL.


As for Statefun (Figure~\ref{fig:concurrency}(c)), in order to achieve maximum throughput, the concurrency level required is significantly higher compared to Orleans. We conjecture several reasons:
(i) Although we aimed for the most performant deployment style, the containerized deployment mode introduces overhead on execution functions due to the virtualization layer;~\footnote{We could not confirm this overhead because we do not find instructions in the documentation to run Statefun in bare metal.}
(ii) The HTTP ingress acts as a single operator, processing requests serially, introducing a bottleneck;~\footnote{We tried using \href{https://nightlies.apache.org/flink/flink-docs-master/api/java/org/apache/flink/streaming/api/functions/source/RichParallelSourceFunction.html}{RichParallelSourceFunction} to enable parallel ingress, but we found that driver requests were arriving out of order in functions, impeding the execution of some transactions.} 
(iii) Although we made sure to adjust the polling rate to a configuration that maximizes Statefun performance, the polling for transaction results coming from the driver to compute metrics in a timely manner invariably introduces a processing overhead. 

Although negligible initially, logging shows a steep increase from the 200 concurrency level, maintaining a stable overhead afterward. The difference in overhead from Orleans lies in the benefits of reentrancy~\cite{reentrancy} found in virtual actors, which minimizes the effect of waiting for responses from external systems.  \newline
\textbf{L\#5.} \textit{The inherent deployment and processing model of the dataflow architecture of Statefun are key factors that harm performance in transactional and event-based workloads.}


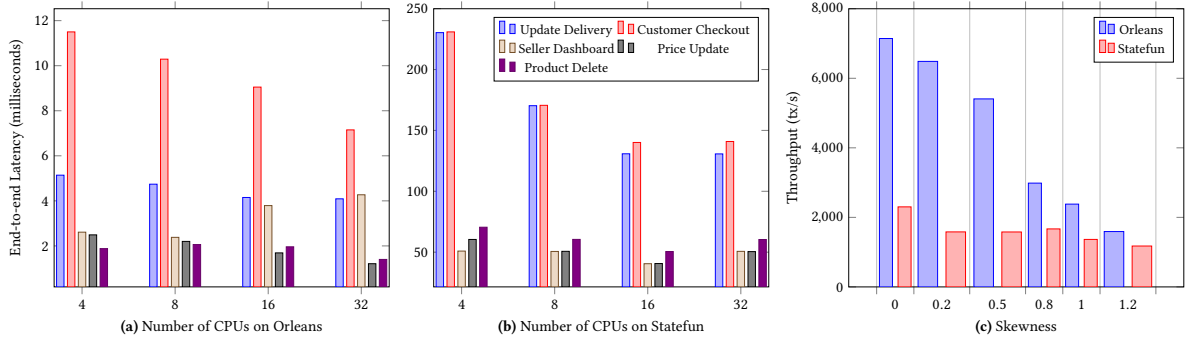
\begin{figure*}
\centering
  \begin{tikzpicture}[scale=0.65]
\begin{axis}[
        ybar=2pt,
        bar width=0.08,
        xlabel={\textbf{(a)} Number of CPUs on Orleans},
        ylabel=End-to-end Latency (milliseconds),
        xtick=data,
        xticklabels={4,8,16,32}, 
        ]
        \addplot
            coordinates {
            (1,5.14)(2,4.74)(3,4.15)(4,4.09)
            };
        \addplot
            coordinates {
            (1,11.50)(2,10.29)(3,9.05)(4,7.15)
            };
        \addplot
            coordinates {
            (1,2.61)(2,2.38)(3,3.79)(4,4.27)
            };
        \addplot
            coordinates {
            (1,2.49)(2,2.20)(3,1.69)(4,1.21)
            };
        \addplot
            coordinates {
            (1,1.88)(2,2.06)(3,1.96)(4,1.40)
            };
    \end{axis}
\end{tikzpicture}
  \begin{tikzpicture}[scale=0.65]
\begin{axis}[
        ybar=2pt,
        bar width=0.08,
        xlabel={\textbf{(b)} Number of CPUs on Statefun},
        xtick=data,
        xticklabels={4,8,16,32},
         legend style={legend columns=2,
        legend pos=north east,font=\small
    }
        ]
        \addplot
            coordinates {
            (1,230.3)(2,170.3)(3,130.8)(4,130.7)
            };
        \addplot
            coordinates {
            (1,230.9)(2,170.6)(3,140.1)(4,140.9)
            };
        \addplot
            coordinates {
            (1,50.9)(2,50.6)(3,40.53)(4,50.67)
            };
        \addplot
            coordinates {
            (1,60.45)(2,50.7)(3,40.67)(4,50.47)
            };
        \addplot
            coordinates {
            (1,70.5)(2,60.5)(3,50.53)(4,60.40)
            };
        \legend{Update Delivery,Customer Checkout,Seller Dashboard,Price Update, Product Delete}
    \end{axis}
\end{tikzpicture}
  \begin{tikzpicture}[scale=0.65]
\begin{axis}[
	x tick label style={
		/pgf/number format/1000 sep=},
	ylabel=Throughput (tx/s),
    xlabel= {\textbf{(c)} Skewness},
    ymin=0, ymax=8000,
    legend style={
        legend pos=north east,font=\small
    },
	ybar interval=0.7,
]
\addplot 
	coordinates {(0,7144) (0.2,6486)
		 (0.5,5407) (0.8,2985) (1,2381) (1.2,1592) (1.5,1284)};
\addplot 
	coordinates {(0,2302) (0.2,1582)
		 (0.5,1580) (0.8,1668) (1,1366) (1.2,1176) (1.5,1016)};
\legend{Orleans,Statefun}
\end{axis}
\end{tikzpicture}
\vspace{-3ex}
\caption{Breakdown Latency (a \& b) and Workload Skewness (c)}
\label{fig:latency}
\vspace{-2ex}
\end{figure*}

\noindent\textbf{Effect of Workload Skew.} 
A skewed workload causes the access of certain records to become more frequent over time. Therefore, it increases contention on a small subset of actors/functions. 
While we measured the platforms under a low skew level in the previous experiments, in this one, we use the \texttt{zipfian} function API in the MathNet library~\cite{mathdotnet} to generate different skewed workloads. To this end, we select different \texttt{zipfian} constants to vary the degree of contention in the workload.

Preliminary experiments showed that fixing seller distribution and varying product skewness showed similar throughput across different skew levels, which led us to investigate the effects of seller skewness. 
Thus, we picked the sellers using a Zipfian distribution while picking products using a uniform distribution. Again, we use 32 CPUs and the concurrency level that maximizes throughput. Figure~\ref{fig:latency}(c) exhibits the Zipfian value of six skew levels used.~\footnote{Statefun crashes on skew levels 1 and 1.2. The metrics are captured up to the crash.}

It is observed that the throughput of Orleans decreases with increasing skewness, following a stable trend. This phenomenon is expected since there is a high contention on certain sellers and their products. Statefun, on the other hand, is less sensitive to workload skewness. In this case, we conjecture that the effects of batching messages play a role, allowing the scheduling of functions not to incur overhead as skewness increases. \newline
\textbf{L\#6.} \textit{A skewed execution introduces a lower performance penalty in Statefun given the inherent batch-oriented execution model.}


\begin{figure}
     \centering
\subfloat[Data Platforms]{\begin{tikzpicture}[scale=0.64]
    \begin{axis}[
        width=8cm,height=3.5cm,
        scale only axis,
        xlabel={Number of CPUs},
        ylabel={Throughput (tx/s)},
        ylabel near ticks,
        xlabel near ticks,
        ymin=0, ymax=8000,
        xtick={0,10,20,30},
        ytick={0,2000,4000,6000,8000},
        legend style={
        legend pos=north west,
        fill=none,
        legend columns=2,
        font=\footnotesize},
        legend cell align={left},
        ymajorgrids=true,
        xmajorgrids=true,
        grid style=dashed
        ]

        \addplot
            coordinates {
            (4,710)(8,1442)(16,3111)(20,4070)(32,7144)
            };
        \addplot
            coordinates {
            (4,543)(8,1304)(16,2463)(20,3011)
            (32,6264)
            };
        \addplot
            coordinates {
            (4,192)(8,473)(16,1145)(20,1786)(32,2302)
            };
        \addplot
            coordinates {
            (4,256)(8,610)(16,1105)(20,1316)(32,1383)
            };
        \addplot
            coordinates {
            (2,72)(4,186)(8,412)(16,630)(24,697)(32,871)
            };
        \addplot
            coordinates {
            (2,66)(4,161)(8,382)(16,534)
            (24,634)(32,577)
            };
            
        \legend{Orleans, Orleans+Logging, Statefun, Statefun+Logging, Orleans TX, Orleans TX+Logging}
    \end{axis}
\end{tikzpicture}}
\par
\subfloat[Composite Solution]{\begin{tikzpicture}[scale=0.64]
    \begin{axis}[
        width=8cm,height=3.5cm,
        scale only axis,
        xlabel={Number of CPUs},
        ylabel={Throughput (tx/s)},
        ylabel near ticks,
        xlabel near ticks,
        ymin=0, ymax=1000,
        xtick={0,10,20,30},
        ytick={0,200,400,600,800,1000},
        legend style={legend pos=south east,font=\footnotesize},
        legend cell align={left},
        ymajorgrids=true,
        xmajorgrids=true,
        grid style=dashed
        ]

        \addplot
            coordinates {
            (2,72)(4,186)(8,412)(16,630)(24,697)(32,871)
            };
           \addplot
            coordinates {
            (2,63)(4,156)(8,314)(16,507)
            (24,679)(32,840)
            };
           \addplot
            coordinates {
            (2,60)(4,142)(8,359)(16,534)
            (24,543)(32,787)
            };
           \addplot
            coordinates {
            (2,58.5)(4,136)(8,360)(16,472)
            (24,524)(32,765)
            };
        \legend{Orleans TX, 
        +View, 
        +View\&Eventual,
        +View\&Causal}
    \end{axis}
\end{tikzpicture}}
     \vspace{-3ex}
     \caption{Scalability}
     \vspace{-5ex}
    \label{fig:scalability}
\end{figure}
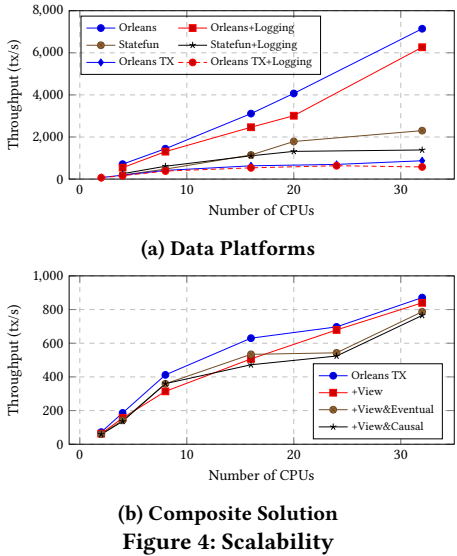

\noindent\textbf{Scalability.}
In this experiment, we evaluate the scalability of both platforms by measuring performance metrics as more resources become available. We do so by increasing the number of cores from 4 to 32 and varying the concurrency level accordingly.

Figure~\ref{fig:scalability}(a) shows Orleans scales linearly as we increase the number of CPUs. In a similar way, Orleans with logging scales nearly linearly. The effects of logging remain low at the start, slightly increasing from 16 CPUs on and remaining stable afterward, matching the expected overhead found in concurrency experiments (Figure~\ref{fig:concurrency}(b)). Statefun can also take advantage of increased computational resources but to a lower degree compared to Orleans. The overhead of logging is nonexistent at the start, and from 16 CPUs on, it impacts Statefun scalability.

On the other hand, we found Orleans Transactions add a significant overhead by enforcing all-or-nothing atomicity and concurrency control. Even runs with a lower concurrency level are dominated by queued lock requests, often leading to message timeouts, thus affecting throughput significantly. We attempted to tune the timeouts of message and lock requests, but no configuration led to improvements over the default one. \newline
\textbf{L\#7.} \textit{Ensuring all-or-nothing atomicity and concurrency control appear as dominant overheads on both platforms.}

Figure~\ref{fig:latency} shows the latency breakdown of the transactions. In line with the throughput, customer checkout, and price update transactions, end-to-end latency in Orleans decreases as more CPUs are available. On the other hand, the seller dashboard shows increasing latency because the more CPUs available, the more concurrent transactions exist in the system, which introduces more customer orders, thus increasing the processing time to compute the query result. Update delivery transactions are not affected by the same phenomena because the processing is independent of the number of orders in progress in the system.



As for Statefun, for customer checkout and update delivery, it is possible to observe latency significantly decreases as more CPUs are available, showing improvements up to 61\% from 4 to 16-32. On the other hand, the overall throughput results are reflected in the longer end-to-end latency.
One of the aspects that drives Statefun's overall latency higher compared to Orleans is inherent to the programming model. Whenever an operation involves interacting with multiple functions, a requester function must send a message to each recipient function and wait for the eventual arrival of the responses. The function must keep track of the responses received through custom-made code, which necessarily involves storing responses in the function's state. On the other hand, in Orleans, multiple actors' calls are encapsulated through promises and do not involve state operations. This scenario occurs in both checkout and update delivery transactions to meet properties 1, 3, 6, and 7. \newline
\textbf{L\#8.}\textit{The absence of native primitives for coordinating asynchronous operations across multiple functions significantly impacts Statefun's overall performance.}

Turning our attention to the composite solution, Figure ~\ref{fig:scalability}(b) shows the overhead of consistent query (+View) and different replication semantics (Eventual and Causal). The results show that the overhead introduced by causal replication is not substantial compared to eventual, which can motivate users to overlook weak semantics in their deployments. On the other hand, the overhead of replication tends to increase as more transactions compete for resources. In the eventual case, we conjecture that, as Orleans Streams are also implemented as actors~\cite{orleans_streams_impl}, they compete with transactional actors for scheduling turns. Similarly, in the causal case, the cost of the increasing number of network I/O operations tends to overcome the benefits of memory-resident data provided by Redis. We omit the results with logging because all run with an average overhead of 5 to 10\%. It is worth noting that we managed to implement the missing criteria adding less than 15\% overhead on average, \textbf{demonstrating that meeting all the benchmark criteria is realistic}. However, the composite solution performance is bounded by Orleans Transactions. \newline 
\textbf{L\#9.} \textit{Offloading replication and continuous queries to external systems inevitably leads to increased network I/O overhead.}
\vspace{-2ex}
\subsubsection{Design Decisions}
\label{subsubsec:design}

In this section, based on the observations and lessons learned identified 
through our case studies,
we devise a set of design decisions 
to advance data platforms and foment new ones. 

In the \textbf{feature} space, we posit the following to address \textbf{L\#1}, \textbf{L\#4}, and \textbf{L\#9}:

\noindent\textbf{D\#1.} \textit{Enhancing microservice platforms with native, system-level support for data replication and querying microservice states is a promising feature.}
This must be paired with performant system-level implementations. The inherent design of existing data platforms may limit the embracement of these missing features, leading to the next two design decisions.

\noindent\textbf{D\#2.} \textit{Integrating external systems and their features into microservice platforms is another potential candidate to relieve the burden on developers.} The case study showed it is realistic to rely on features provided by external systems. However, fulfilling missing requirements must not jeopardize performance.

\noindent\textbf{D\#3.} \textit{Integrating the features from external systems must be accompanied by optimizations to hide latencies associated with cross-system coordination to prevent substantial performance penalty.}

In terms of \textbf{performance}, lessons \textbf{L\#5} and \textbf{L\#8} demonstrate the values of the following design decisions:


\noindent\textbf{D\#4.} \textit{The event processing model and task scheduling in microservice platforms
should reduce the impact of the
long latency of certain multi-microservice workflows (e.g., customer checkout) on shorter transactions (e.g., price updates).}
These facilities should ideally be transparent to developers, as in Orleans. In addition, designing primitives to 
circumvent concurrency limitations 
inherent to the programming model is another opportunity, as follows.\newline
\noindent\textbf{D\#5.} \textit{Multiplexing multiple non-conflicting transactions within a scheduling unit (e.g., enabling grain reentrancy in Orleans) is a promising technique to speed up microservice workloads,} enhancing throughput by running non-conflicting transactions concurrently. 


From a \textbf{system architecture} perspective, lessons \textbf{L\#2} and \textbf{L\#6} derive the following:

\noindent\textbf{D\#6.} \textit{Holistically managing state updates and message queue operations found in Statefun appears to be a promising approach for ensuring exactly-once semantics}. That relieves developers from writing idempotent code to prevent duplicate executions, which may not always be possible. These should be paired with efficient solutions to schedule application functions.

\noindent\textbf{D\#7.} \textit{Batching multiple conflicting transactions is an effective performance technique for highly-contended microservice workloads.}
Statefun inherits the operator execution model of Flink, which processes event streams in batches. Orleans, on the other hand, tracks message timeouts individually, which limits batching opportunities.

Lesson \textbf{L\#7} derives better \textbf{algorithmic design} and lesson \textbf{L\#3} addresses the usefulness of proper \textbf{abstractions}, respectively:

\noindent\textbf{D\#8.} \textit{Devising new algorithms and techniques for achieving all-or-nothing atomicity and concurrency control is critical for microservice performance.} Recent research shows it is possible to speed up Orleans Transactions by 4X through deterministic scheduling~\cite{snapper}. Offering weaker isolation levels that can still achieve the required properties is another promising direction. For instance, customer statistics can be updated in any order as long as the events are processed exactly once.

\noindent\textbf{D\#9.} \textit{Virtualized representation of application functions and associated states, such as through a virtual actor in Orleans, appears as an effective abstraction} to manage deployment, discovery, distribution, scheduling, and failures holistically in microservices.

\section{Concluding Remarks}


Our experiments with five implementations of \OM\ in three competing systems have shown varied results under different workloads, showing the ability of \OM\ to pinpoint missing core data management features and the impact of the programming abstractions and architectural design of these systems on performance. 
The experiments also highlight the overhead of all-or-nothing atomicity, concurrency control, constraint enforcement, replication semantics, consistent queries, audit logging, and asynchronous processing of events. \textbf{It is worth noting these cannot be obtained by existing microservice benchmarks ($\S$~\ref{sec:background})}.
Furthermore, from the lessons learned obtained through \OM, we derive design decisions to advance the development of data systems, highlighting that \OM\ will support designing futuristic data management systems for microservices.

\begin{acks}
This work was partially supported by Independent Research Fund Denmark under Grant 9041-00368B.
All of the computations for this project were performed on the UCloud interactive HPC system.

\noindent Dedicated to the memory of José Apolinário Nunes $\dagger$.
\end{acks}



\bibliographystyle{ACM-Reference-Format}
\bibliography{main}


\end{document}